\documentclass{emulateapj}

\usepackage{multirow}

\newcommand{\boom}{{\sc Boom\-erang }}

\newcommand{\boomn}{{\sc Boom\-erang}}
\newcommand{\cmb}{{\sc cmb }}
\newcommand{\cmbn}{{\sc cmb}}
\newcommand{\te}{$\langle TE \rangle$ }
\newcommand{\ten}{$\langle TE \rangle$}
\newcommand{\tb}{$\langle TB \rangle$ }
\newcommand{\tbn}{$\langle TB \rangle$}

\newcommand{\TT}{$\langle TT \rangle$ }
\newcommand{\TTn}{$\langle TT \rangle$}
\newcommand{\linka}{http://oberon.roma1.infn.it/boomerang/b2k}
\newcommand{\linkb}{http://cmb.phys.cwru.edu/boomerang}
\def\p{{\bf p}}
\def\pML{{\bf p_{\rm ML}}}

\slugcomment{To appear in  The Astrophysical Journal}

\shorttitle{\te of \cmb from \boomn}
\shortauthors{Piacentini {\it et al.}}

\begin{document}
  
  
  \title{A measurement of the 
polarization-temperature \\ angular cross power spectrum of the 
    Cosmic Microwave Background \\from the 2003 flight of BOOMERANG}
  

\author{
F.~Piacentini\altaffilmark{1},  
P.A.R.~Ade\altaffilmark{3}, 
J.J.~Bock\altaffilmark{4,14},
J.R.~Bond\altaffilmark{5},
J.~Borrill\altaffilmark{6,17},
A.~Boscaleri\altaffilmark{7},
P.~Cabella\altaffilmark{16},
C.R.~Contaldi\altaffilmark{5,15},
B.P.~Crill\altaffilmark{8},
P.~de~Bernardis\altaffilmark{1},
G.~De~Gasperis\altaffilmark{16},
A.~de~Oliveira-Costa\altaffilmark{12},
G.~De~Troia\altaffilmark{1},
G.~Di~Stefano\altaffilmark{11},
E.~Hivon\altaffilmark{8},
A.H.~Jaffe\altaffilmark{15},
T.S.~Kisner\altaffilmark{2,9},
W.C.~Jones\altaffilmark{14},
A.E.~Lange\altaffilmark{14},
S.~Masi\altaffilmark{1},
P.D.~Mauskopf\altaffilmark{3},
C.J.~MacTavish\altaffilmark{13},
A.~Melchiorri\altaffilmark{1,18},
T.E.~Montroy\altaffilmark{2},
P.~Natoli\altaffilmark{16,19},
C.B.~Netterfield\altaffilmark{13},
E.~Pascale\altaffilmark{13},
D.~Pogosyan\altaffilmark{20},
G.~Polenta\altaffilmark{1},
S.~Prunet\altaffilmark{10},
S.~Ricciardi\altaffilmark{1},
G.~Romeo\altaffilmark{11},
J.E.~Ruhl\altaffilmark{2},
P. Santini\altaffilmark{1},
M.~Tegmark\altaffilmark{12},
M. Veneziani\altaffilmark{1}, and
N.~Vittorio\altaffilmark{16,19}.
}

\affil{
$^1$ Dipartimento di Fisica, Universit\`a di Roma La
Sapienza, Roma, Italy \\
$^2$ Physics Department, Case Western Reserve University,
		Cleveland, OH, USA\\
$^{3}$ Dept. of Physics and Astronomy, Cardiff University, 
		Cardiff CF24 3YB, Wales, UK \\
$^4$ Jet Propulsion Laboratory, Pasadena, CA, USA\\
$^5$ Canadian Institute for Theoretical Astrophysics, 
		University of Toronto, Toronto, Ontario, Canada\\
$^6$ Computational Research Division, Lawrence Berkeley National Laboratory, Berkeley, CA, USA\\
$^7$ IFAC-CNR, Firenze, Italy\\
$^8$ IPAC, California Institute of Technology, Pasadena, CA, USA\\
$^{9}$ Dept. of Physics, University of California, Santa Barbara, CA, USA\\
$^{10}$ Institut d'Astrophysique, Paris, France\\
$^{11}$ Istituto Nazionale di Geofisicae Vulcanologia, Roma,~Italy\\
$^{12}$ Dept. of Physics, Massachusetts Institute of Technology, Cambridge,  MA, USA\\
$^{13}$ Physics Department, University of Toronto, Toronto, Ontario, Canada\\
$^{14}$ Observational Cosmology, California Institute of
Technology, Pasadena, CA, USA\\
$^{15}$ Theoretical Physics Group, Imperial College, London, UK\\
$^{16}$ Dipartimento di Fisica, Universit\`a di Roma Tor
Vergata, Roma, Italy\\
$^{17}$ Space Sciences Laboratory, UC Berkeley, CA, USA\\
$^{18}$ INFN, Sezione di Roma 1, Roma, Italy\\
$^{19}$ INFN, Sezione di Roma 2, Roma, Italy\\
$~{20}$ Department of Physics, University of Alberta, Edmonton, AB, Canada\\
}



\begin{abstract}

We present a measurement of the 
temperature-polarization angular cross power
spectrum, \ten, of the Cosmic Microwave Background.
The result is based on $\sim 200$ hours of data from 8 polarization
sensitive bolometers operating at 145~GHz  during the 2003
flight of \boomn.  We detect a significant \te correlation in
the $\ell$-range between 50 and 950 with a statistical 
significance $> 3.5~\sigma$.
Contamination by polarized foreground emission and systematic effects
are negligible in comparison with statistical uncertainty.
The  spectrum is consistent with previous detections and with the
``concordance model'' that assumes adiabatic initial conditions.  This
is the first measurement of \te using bolometric detectors.

\end{abstract}

\keywords{cosmology: cosmic microwave background}

\section{Introduction}
Since Rees pioneering work \citep{rees}, polarization of the
Cosmic Microwave Background (\cmbn) 
has been at the center
of several theoretical studies.  

Detailed numerical 
predictions have been made in the framework of 
standard inflationary models with primordial adiabatic and
scale-invariant fluctuations \citep[see e.g.][]{be84,sz}. 
\cmb polarization data is highly useful for cosmology since
it can shed light, for example, on the process of reionization 
of the intergalactic medium \citep[see e.g.][]{efs88,zald96}, on the
amplitude of the inflationary gravity waves background 
\citep{crittenden,sz96,kam96} and on the nature of
primordial perturbations \citep[see e.g.][]{spergel97,kavi00}.
Moreover, polarization can provide further evidence for coherent acoustic
oscillations in the early universe, since in this case peaks in the 
temperature and polarization power spectra are expected to be 
180 degrees out of phase \citep{koso98}. Unfortunately, 
given the small amplitude of the signal,
current \cmb polarization data, while providing an important
confirmation of the standard scenario, are unable to provide
useful constraints on the parameters of the model. As first
suggested in \cite{coulson94}, measuring the temperature-polarization cross
correlation is easier, since the signal is higher  
and carries most of the cosmological
information present in the polarization data.

Previous detections of the 
temperature-polarization angular cross power spectrum have
been obtained by the WMAP satellite~\citep{kogut03} and by 
the DASI \citep{dasi1,dasi2} and CBI \citep{cbi2004} interferometers.

In this paper we present new observations of
the temperature-polarization cross power spectrum 
of the Cosmic Microwave Background anisotropy 
obtained by the \boom experiment flown in Jan. 2003 (B03).
Results on the temperature and polarization 
power spectra alone are presented in two companion papers
\citep{jones05,montroy05}; the instrument and the analysis pipeline
producing the maps of temperature and polarization are
described in~\cite{masi05}; the cosmological parameter
extraction is in~\cite{mactavish05}.

In the present paper we will follow the notation of \cite{sz96}
\citep[but see also][]{kam96} in which polarization is expressed as two 
linear combinations of spin$\pm 2$ multipole moments which have
opposite parities, the so-called 
$E$ (electric) and $B$ (magnetic) modes. 
In standard cosmological models, the magnetic-type parity
combination does not cross-correlate with temperature or the electric-type
parity combination. The cosmological information in the
polarization-temperature correlation is therefore present only in the
\te angular power spectrum. In this paper we will also present
constraints on $\langle TB \rangle$ as useful check for systematics
and foregrounds.
Non zero $\langle TB \rangle$ may also appear in exotic theories 
due for example to the presence of helical flows in the primordial
plasma at the time of recombination \citep{pogosian01}.

We have performed the analysis of the B03 data using two
completely independent pipelines, with 
different procedures for the pointing solution, 
data cleaning, deconvolution, map-making and noise estimation,
and different estimation of the calibration factors, beams, receivers
transfer functions, polarization efficiencies and polarizer
angles. One pipeline was developed in Italy (IT), the other in 
North America (NA). 
Pipelines details are in \cite{masi05}
and will be described furthermore in 
subsequent papers.
The most important result from this splitting
is the overall agreement, which enhances confidence in the
result. A comparison of the result from the two pipelines allows a
measure of the sensitivity of the result to details of the analysis.

\section{\te estimation}
We use data from 8 channels at 145 GHz, composed of 4 pairs 
of Polarization Sensitive Bolometers (PSB)- (W1, W2), (X1, X2), (Y1,Y2)
and (Z1, Z2), with effective angular resolution of 11.5~arcminutes 
(full width half maximum, including pointing jitter).
Performance and characteristics of those devices are
in~\cite{masi05}, together with the full description
of the instrument and of the temperature and polarization
maps that are used for the analysis presented here.

With polarization sensitive bolometers \boom produces maps of 
the three Stokes parameters, $I$, that describe fluctuations
in the brightness of the radiation, $Q$ and $U$ that describe 
the linear polarization.
The intensity of the \cmb is conveniently described in terms of 
temperature fluctuations $\Delta T$ of a black-body respect to
a 2.725~Kelvin black-body,  
and can be decomposed in spherical harmonics as
$\Delta T(\hat n) = \sum_{\ell m}a^T_{\ell m} Y_{\ell m}(\hat n) $.
Similarly the linear polarization  $Q+iU$ is decomposed  
using the spin-2 weighted basis $_{\pm 2}Y_{\ell m}$
\begin{equation}
(Q \pm i U)(\hat n) = \sum_{\ell m} \left(a_{\ell m}^E \mp i a_{\ell m}^B \right)  ~_{\pm 2}Y_{\ell m}(\hat n) 
\end{equation}
thus defining the scalar field
$E(\hat n) = \sum_{\ell m}a^E_{\ell m} Y_{\ell m}(\hat n)$ 
and the pseudo-scalar
$B(\hat n) = \sum_{\ell m}a^B_{\ell m} Y_{\ell m}(\hat n)$.
In the hypothesis that those quantities are Gaussian 
distributed and that the early Universe is isotropic, 
the cosmological information is 
encoded in the standard deviations and correlations
of the coefficients:
\begin{equation}
\langle XY \rangle = \langle  a^{X*}_{\ell m} a^Y_{\ell'm'} \rangle = C_\ell^{XY} \delta_{\ell \ell'}\delta_{mm'}
\end{equation}
where the pairs 
$\langle XY \rangle$ can be 
\TTn, $\langle EE \rangle$, $\langle BB \rangle$, \ten, \tb and $\langle EB \rangle$.
Given the isotropy, these power spectra 
can be estimated by averaging over $m$ at each multipole 
number $\ell$.

Both IT and NA power spectra estimation pipelines are based on 
the MASTER method \citep{master} that computes the
pseudo-$a_{\ell m}$
on a fraction of the sphere defined by the function
$W(\hat n)$ that takes into account  weighting and sky coverage. 
This yields the definition of mode-mode coupling kernels
that depend only on the weighted scheme.
Using an appropriate $\ell$-binning it is possible to solve
for the underlying angular power spectra, taking into account 
the binning operator, the angular resolution of the instrument, the 
pixelization, and the filtering of time stream.
The quantity that is normally used for the binning
is the flattened power spectrum  
$\mathcal C_\ell = \ell(\ell+1) C_\ell /2\pi$.
For a set of $n$ bins indexed by $b$,
with boundaries $\ell_{\rm{low}}^{(b)} <\ell_{\rm{high}}^{(b)}<\ell_{\rm{low}}^{(b+1)} $,
the binning operator is defined as
\begin{equation}
\label{eqn:binning}
P_{\ell}^b = 
\cases 
{
\frac{1}{\ell_{\rm{low}}^{b+1}-{\ell_{\rm{low}}^{b}}} & 
if $2 \leq \ell_{\rm{low}}^{(b)} \leq \ell \leq \ell_{\rm{low}}^{(b+1)}$\cr
0 & otherwise.\cr
}
\end{equation}
and the power in each bin (hereafter band powers) are 
$\mathcal C_b = P_{\ell}^b \mathcal C_\ell$.

The method is based on Monte Carlo simulations of signal-only
time-streams, from simulations of the \cmbn, and of noise-only time-streams, 
from simulations of the instrument. Both simulated
data-streams are processed in the
same way as the real data, in order to take into account the 
overall
effect of data filtering and partial sky coverage, and to estimate the 
noise bias to be removed in the power spectra estimation. 

The 
signal simulations are obtained from random realization of the
\cmb sky, in temperature and polarization, given an
underlying cosmological model, projected in 
a time stream according to the \boom pointing solution.
The noise simulations 
are obtained from random realization of the noise power
spectrum, iteratively estimated,
taking into account noise correlation between
channels as described in \citet{masi05}.

The covariance matrix ${\bf M}_{bb'}$, that defines the
uncertainties in the $\mathcal C_b$ determination, is estimated by 
Monte Carlo simulations of signal plus noise as in~\cite{master}.
An approximation of the diagonal part of this matrix is given by
\begin{eqnarray}\label{eqn:noise}
\sigma^2_{TE,b} =&& {{2}\over{(2 \ell_b +1) f_{\rm{sky}} \Delta \ell}}\\
\nonumber
&&\left [
\mathcal C^2_{TE,b} + 
\left( \mathcal C_{T,b} + {{\mathcal N_{T,b}}\over { B_b}^2} \right )
\left( \mathcal C_{E,b} + {{\mathcal N_{E,b}}\over { B_b}^2} \right )
\right ]
\end{eqnarray}
where $f_{\rm{sky}}$ is the effective observed fraction of sky,
$\mathcal C_{T,b}$, $\mathcal C_{E,b}$ and  $\mathcal C_{TE,b}$
are the band powers of the temperature, the polarization and
the polarization-temperature correlation respectively, 
$\mathcal N_{T,b}$ and  $\mathcal N_{E,b}$
are the band powers of the noise
in the temperature map and in the $E$ map,
$B_b$ is the spherical harmonic transform of the
beam, and $\Delta \ell$ is the bin width.

\subsection{Weighting}
The last flight of \boomn, described in~\cite{masi05},
was split into three parts: a deep observation over 0.22\% 
of the sky, centered at 
$RA = 82.5^\circ$, $DEC=-45^\circ$ (hereafter deep region),
a shallow observation on a region of covering the 1.8\%
of the sky (including the deep region)
centered in the same coordinates (hereafter shallow region), and
observation 
of the Galactic plane that is not used in these power spectra analysis.
Wide coverage and deep integration are both important 
for the quality of the result. The wide coverage of the shallow 
region is useful to reduce sample variance, the deep integration of the deep
region to reduces the statistical noise.

The two pipelines use different methods to combine the 
data to obtain a compromise of sample variance and noise.
In the NA pipeline we perform 
independent analysis of the shallow scans and of the deep scans,
computing the respective $a_{\ell m}^T$ and $a_{\ell m}^E$.
We then estimate four \te cross spectra, 
$\langle T_s E_s \rangle$, 
$\langle T_s E_d \rangle$, 
$\langle T_d E_s \rangle$, 
$\langle T_d E_d \rangle$, 
with the
relative correlation matrices and combine the 
spectra appropriately (C.R.~Contaldi, in preparation).
In the IT pipeline, a single map with all the scans
is used. The $a_{\ell m}^T$ are computed on the shallow
region, the $a_{\ell m}^E$ on the deep. 
The effect of such a double coverage is taken into
account in the transfer function and kernel used
to derive the \te spectrum.

\subsection{Result and significance}

\begin{deluxetable*}{c|cccc|cccc}
\tablecolumns{9} 
\tablewidth{0pc} 
\tablecaption{
\boomn-{\sc 03}
Temperature-polarization cross power spectra
band powers.
\label{tab:te} }
\tablehead{
~&
\multicolumn{4}{c}{NA} &
\multicolumn{4}{c}{IT} \\
\colhead{ $\ell_b$ } & 
\colhead{ $\mathcal C_b^{TE}  $ }           &
\colhead{ $\Delta \mathcal C_b^{TE}  $ }    &
\colhead{ $\mathcal C_b^{TB}  $ }           &
\colhead{ $\Delta \mathcal C_b^{TB}  $ }    &
\colhead{ $\mathcal C_b^{TE}  $ }           &
\colhead{ $\Delta \mathcal C_b^{TE}  $ }    &
\colhead{ $\mathcal C_b^{TB}  $ }           &
\colhead{ $\Delta \mathcal C_b^{TB}  $ }
}
\startdata
        51&          22&          32&           2&          33 &          -9&          51&           2&          51\\
       150&         -51&          27&         -18&          27&         -71&          39&         -68&          37\\
         250&          40&          32&          -9&          31&         125&          46&         -73&          42\\
         350&          58&          28&         -16&          27&          63&          34&          29&          30\\
         450&         -90&          29&           7&          28&         -69&          40&         -17&          36\\
         550&          40&          39&          -2&          35&          20&          44&          66&          39\\
         650&         -18&          45&          -2&          42&          -8&          65&          91&          60\\
         750&         -86&          60&         -88&          58&         -88&          56&          65&          51\\
         850&          62&          74&          74&          72&          77&          86&          69&          82\\
         950&         -61&          90&         -70&          88&        -115&          72&         199&          65\\
        1500&          48&          81&        -133&          81&          90&         105&         -12&         100\\
\enddata
\tablenotetext{}{
Units are $\mu K_{CMB}^2$. The uncertainties are given by the
square root of the diagonal part of the covariance matrices.
The first and the last bins must be excluded from 
any analysis since can be contaminated by
instrumental effects. 
Complete results, including window functions and covariance matrices,
are available at \linka and \linkb.
}
\end{deluxetable*}

\begin{figure}[tb]
\plotone{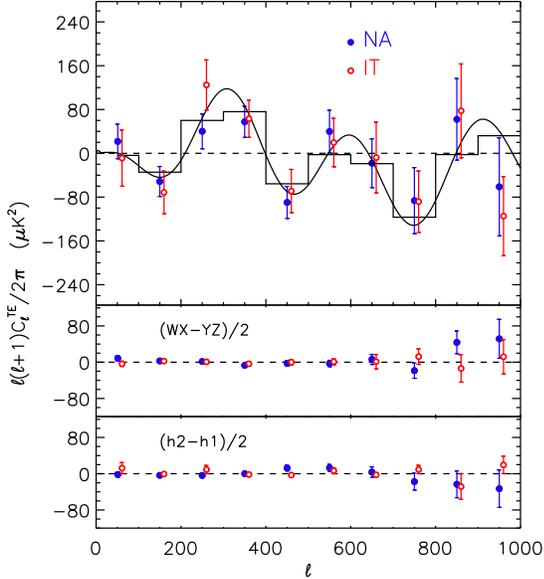}
\caption{The \te power spectrum band powers
for the NA (filled circles) and IT (open circles) pipelines.
The upper part of the plot reports data with errorbars, the
fiducial model 
($\Lambda$CDM model fit to
WMAP (year~1), Acbar, and CBI)
as a black curve and the binned fiducial model 
as histogram. 
The middle and bottom plots are the results of two different consistency  
tests, obtained splitting the data in channels 
(WX-YZ)
and in time (half 2 - half 1) respectively.
In the low-$\ell$ part of the plot is evident the effect of 
a different weighting scheme between IT and NA, while at large multipoles
the result is dominated by the same instrumental performances.}
\label{fig:te}
\end{figure}

\begin{figure}[tb]
\plotone{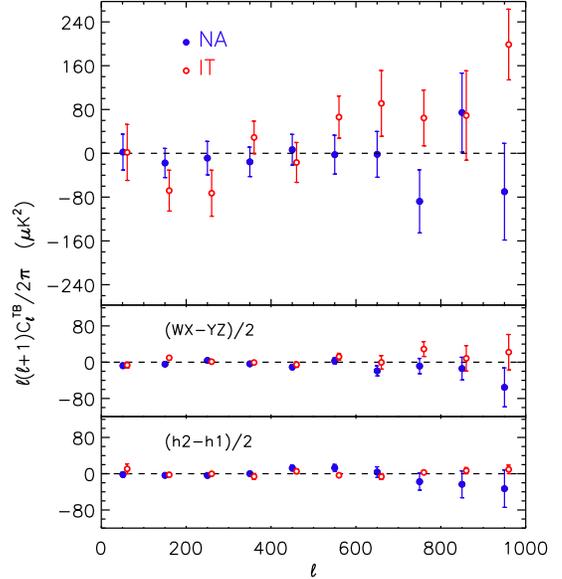}
\caption{The \tb power spectrum band powers
for the NA (filled circles) and IT (open circles) pipelines.
The upper part of the plot reports the \tb data with error-bars. 
The middle and
bottom plots are the results of two different consistency  
tests, obtained splitting the data in channels (WX-YZ)
and in time (half 2 - half 1) respectively.
}
\label{fig:tb}
\end{figure}
 
\begin{figure}[tb]
\plotone{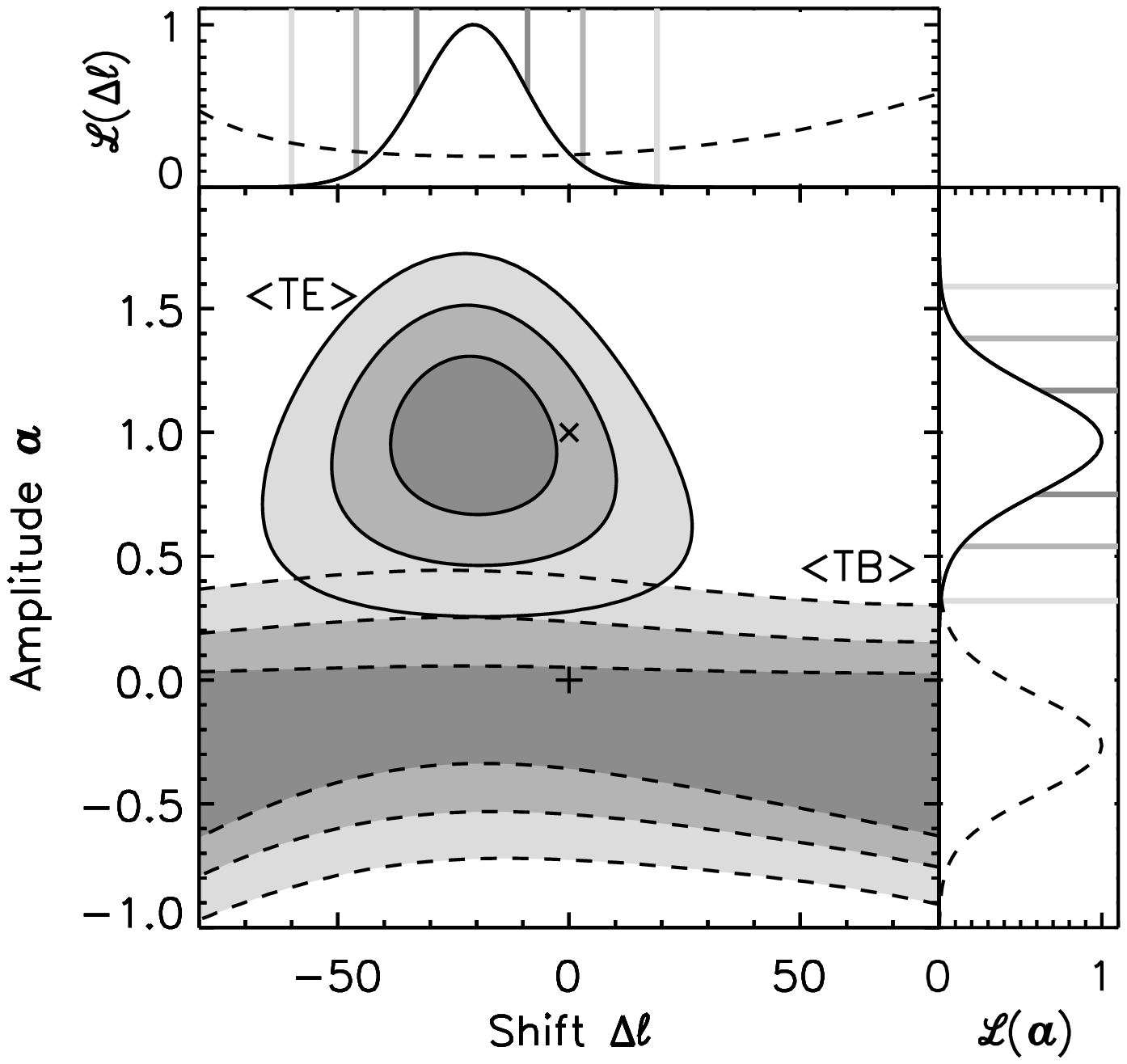}
\caption{
Likelihood of the parameters $a$ and $\Delta \ell$.
Parameter $a$ is defined as the amplitude of 
the \te fiducial model, $\Delta \ell$ is the shift
in multipole $\ell$ applied to the \te fiducial model.
The continuum lines are for the \boom \te data, the
dashed lines are for \tb data, compared to the same
\te fiducial model.
In the central plot is reported the two dimensional 
likelihood, $\mathcal L(\Delta \ell, a)$; the contours
are 1, 2, and 3 $\sigma$, corresponding to 68.3, 95.4 and
99.7\% of probability. The $\times$ symbol is the expected 
value for \te given the fiducial model, the {\bf +} symbol
is the expected value for \tbn. 
In the right plot is reported the 
$\mathcal L (a)$ marginalizing over $\Delta \ell$, and in the top
plot, the $\mathcal L (\Delta \ell)$ marginalizing over $a$.
In the right and top plot, the gray lines are
1, 2, and 3 $\sigma$ boundaries for the \te data.
The \tb data likelihood is used to test the  presence of foregrounds and
systematic effects that would affect \te and \tb in the same
way. This plot is obtained using the IT dataset with 
a binning width of 50 multipole numbers. 
The NA dataset gives a similar result. 
}
\label{fig:significance}
\end{figure}
 
\begin{deluxetable*}{rlccccc}
\tablecolumns{7} 
\tablewidth{0pc}
\tablecaption{\te and \tb statistics
\label{tab:pte}} 
\tablehead{\colhead{~} &
\colhead{Test} & 
\colhead{d.o.f.} &
\colhead{$\chi^2$}& 
\colhead{$\Lambda$} & 
\colhead{PTE} & 
\colhead{$\sigma_{\mbox{reject.}}$}}
\startdata
\multirow{4}{*}{NA} 
&\te compared to fiducial & 9 & 5.81&  0.48   & {0.62} &  ~ \\
&\te compared to zero     & 9 & 23.1&  8.23   & {$2.6 \times 10^{-4}$} & 3.5 \\
&\tb compared to zero     & 9 & 4.94&  0.08   & {0.92} &  ~ \\
&\tb compared to fiducial \te & 9 & 24.9& 10.5   & {$2.9 \times 10^{-5}$} & 4.0\\
\hline
\multirow{4}{*}{IT}
&\te compared to fiducial & 9 & 4.59&  1.83   & {0.26} &  ~ \\
&\te compared to zero     & 9 & 20.8&  11.3   & {$1.2 \times 10^{-5}$} & 4.2 \\
&\tb compared to zero     & 9 & 15.1&  1.86   & {0.16} &  ~ \\
&\tb compared to fiducial \te & 9 & 34.5 & 20.7   & {$1.0 \times 10^{-9}$} & 6.0\\
\enddata
\tablecomments{Significance of the \te and \tb results respect to models.
The first bin is not used in this analysis
since it can be contaminated by instrumental effects.
}
\end{deluxetable*}

The results for the \te and \tb power spectra are reported in 
the upper panel of Figures~\ref{fig:te} and~\ref{fig:tb} respectively
and in Table~\ref{tab:te}.

Both pipelines assume a flat shape for the power in each band
(see equation~\ref{eqn:binning}).
For comparison with model band powers, the IT pipeline assumes flat band
power window functions while the NA pipeline computes from 
the {\sc Xfaster} fisher matrix estimator
the band power window functions
$\mathcal W_\ell^b$, that are used, in place of $P_\ell^b$, 
to convert a model power spectra 
$\mathcal C_\ell^{\rm{mod}} = \ell(\ell + 1 ) C_\ell^{\rm{mod}}/2\pi$ 
into the theoretical band powers as  
\begin{equation}
\mathcal C_b^{\rm{mod}} = \frac{\mathcal I \left [ \mathcal W_\ell^b \mathcal C_\ell^{\rm{mod}}\right ]}
{\mathcal I \left [ \mathcal W_\ell^b \right ]}
\end{equation}
where 
$\mathcal I\left[f_\ell \right] = \sum _\ell \frac{\ell + \frac 12}{\ell (\ell +1)}f_\ell$
is the logarithmic integral defined in \cite{bjk2000}.
Band powers, covariance matrices 
and window functions are available at 
the \boom web-pages\footnote{{\tt \linka \\ \linkb}}.

To quantify the agreement  of the detection with standard cosmology,
we compare the
result to  $\mathcal C_b^{\rm{mod}}$, the theoretical band powers of
a fiducial model
given by the $\Lambda$CDM model of~\cite{spergel03} fit to
WMAP (year~1), Acbar, and CBI, which we scale by a factor $a$ and shift by
$\Delta\ell$.
We compute the two dimensional likelihood as
\begin{eqnarray}
&&\mathcal{L}(a,\Delta \ell) \propto\\
\nonumber
&&\exp \left(-{1 \over 2} \sum_{bb'} 
\left (\mathcal C_b-a \cdot \mathcal C_{b,\Delta \ell}^{\rm{mod}}\right ) 
{\bf M}_{bb'}^{-1}
\left(\mathcal C_{b'}-a \cdot \mathcal C_{b',\Delta \ell }^{\rm{mod}} \right) \right)
\end{eqnarray}
where $\mathcal C_{b',\Delta \ell }$ 
are the band powers after shifting by $\Delta \ell$
the power spectrum. Given the fact that the
\te power spectrum crosses the zero several times, to improve the
detection 
we used in this analysis a binning width of 50 multipole numbers, and the corresponding
covariance matrix.

The result is reported in Figure~\ref{fig:significance}, 
together with the one-dimensional likelihoods obtained
by marginalization.
For the \te data, the likelihood defined as above favours
a multipole shift in the range $-46 < \Delta \ell < 3$ 
($-38 < \Delta \ell < 20$ for NA)
and an amplitude in the range $0.54<a<1.38$ 
($0.40<a<1.30$ for NA) 
at 95\% of
probability. The data are thus in agreement with the 
amplitude and phase of the \te power spectrum predicted
from the \TT power spectrum under the
hypothesis of adiabatic initial perturbations.
For the \tb data, the likelihood (with respect to a \te fiducial
model) does not constrain the multipole shift and 
gives an amplitude in the 
$-0.67<a<0.13$
($-0.43<a<0.45$ for NA) 
range at 95\% of probability.

To compare our data to a model $\mathcal H_0$ characterized by 
a parameter set ${\bf p}$, 
we define the goodness of fit of the model as 
\begin{equation}
\Lambda(\p) = \log \left( {{\mathcal L(\pML)}
\over {\mathcal L(\p)}} \right)
\end{equation}
where $\mathcal L(\pML)$ and $\mathcal L(\p)$ are the values of the 
likelihood at the maximum and for the parameters $\p$ of the model.
In the approximation that the likelihood function $\mathcal L(\p)$ is
multivariate Gaussian near its peak, 
the goodness of fit reduces to $\Lambda = \Delta \chi^2/2$ and the 
probability of total exclusion is defined by the incomplete
Gamma function
\begin{equation}
PTE(\Lambda) ={{1}\over{\Gamma(N/2)}} \int_\Lambda^\infty e^{-x} x^{N/2-1} dx 
\end{equation}
where $N$ is the number of parameters in the model,
which is 2, $a$ and $\Delta \ell$ in out case.
A set of tests performed using 9 bins between $\ell = 100$ and
$\ell = 1000$ is reported in Table~\ref{tab:pte}.
In that Table, $\sigma$ represents
the number of standard deviations of
a Normal distribution to have the same $PTE$ as $\Lambda$ does.
The \te = 0 model is rejected at 3.5~$\sigma$ (4.2~$\sigma$ for IT) and the 
$\langle TB \rangle =\langle TE \rangle^{th}$ is rejected at 
4.1~$\sigma$ (6.0~$\sigma$ for IT). 
The complete results of consistency with cosmological models
and parameter extraction treatment
is reported in \cite{mactavish05}.

\section{Control of foregrounds}
Polarization generated by foregrounds presents no 
global symmetry and thus is  expected to contaminate both 
$E$ and $B$ components of the \cmb in a similar 
way~\citep[see e.g.][]{teg_fore2000,tucci_fore2002,bacci_fore2001}.
The \tb power spectrum can be used to test such a 
contamination. The \tbn=0 result presented in Table~\ref{tab:pte}
is the main evidence that the \te result is not contaminated.

\begin{figure}[tb]
\epsscale{1}
\plotone{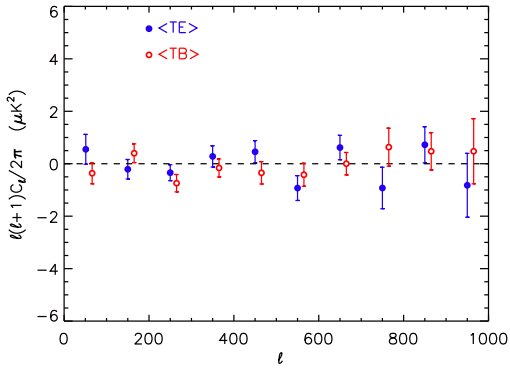} 
\caption{
Dust contamination. 
Filled circles are $\langle T_{dust} E_{B03} \rangle$,
open circles
$\langle T_{dust} B_{B03} \rangle$.
The dust contamination to \te is two order of
magnitude lower than the measured \ten. The B03 data 
are from the IT pipeline.\\
\label{fig:schb2k}}
\end{figure}

Moreover we can directly test the contamination due
to dust by a correlation of our data with a dust map.
If we assume that the temperature and polarization 
seen by \boom are
a superposition of \cmb and dust, $T_{B03} = T_{CMB}+ T_{dust}$
and $E_{B03} = E_{CMB}+ E_{dust}$ (and the same for $B$), then 
\begin{equation}
\langle TE \rangle_{B03} = \langle T_{CMB} E_{CMB} \rangle
+ \langle T_{dust} E_{dust} \rangle
\end{equation}
where we assume that dust and \cmb are not correlated
($\langle T_{CMB} E_{dust} \rangle = \langle T_{dust} E_{CMB} \rangle =0$).
Under the same assumption, the contaminating term 
$\langle T_{dust} E_{dust} \rangle$ can be estimated
by 
\begin{equation}
\langle T_{dust} E_{dust} \rangle \simeq \langle T_{dust} E_{B03} \rangle
\end{equation}
We estimate $T_{dust}$  by using
IRAS maps re-calibrated with DIRBE at 100~$\mu m$, 
extrapolated to our 
wavelength with model 8 in~\cite{fink99}
as described in~\cite{masidust}.
We resample the extrapolated dust map with 
the \boom scan strategy,  and then recreate a dust map with 
the same time domain filtering, flagging, and map-making 
algorithm as the \boom map.
As shown in Figure~\ref{fig:schb2k} this contaminant is
compatible with zero and two orders of magnitude lower than
the detected \ten.

\section{Control of systematic effects}
The standard test to detect systematic effects consists 
of splitting the data in two subsets (jackknife),
making a differenced map using the two subsets
and calculating the power spectra 
of the differenced map divided by two to maintain 
the same noise statistics as an average map.
The result must be 
consistent with zero. This test is particularly effective 
because the sample variance goes to zero and the 
noise in equation (\ref{eqn:noise}) reduces to
\begin{equation}
\sigma^2_{TE,b} = {{2}\over{(2 \ell_b +1) f_{\rm{sky}} \Delta \ell}}
\left (
{{\mathcal N_{T,b} \mathcal N_{E,b} }\over { B_b^4}} \right )
\end{equation}
We performed two such tests, splitting the data in 
time and in channels. As our temporal splitting we
take the first half of the scans on the shallow region
and the first half of the scans on the deep region 
versus the second half of shallow plus second half of
deep. As our channel splitting we take two PSB pairs
(W1, W2, X1, X2) 
versus the other two PSB pairs
\citep[Y1, Y2, Z1, Z2 for focal plane description see][]{masi05}. 
Results are 
reported in the bottom panels of Figures~\ref{fig:te}
and~\ref{fig:tb} and in Table~\ref{tab:chi2}, showing 
remarkable consistency with zero.
This, along with the \tbn=0 result presented above,
gives strong evidence that the dataset is free from
significant systematics.

\begin{deluxetable}{rlccc}
\tablecolumns{5}
\tablecaption{Consistency tests \label{tab:chi2}  }
\tablewidth{0pc} 
\tablehead{\colhead{~}&\colhead{Test} &\colhead{d.o.f.} &\colhead{$\chi^2$} & \colhead{$P_>$}}
\startdata
\multirow{4}{*}{NA} 
&\te temporal  & 9  & 16.1 & 0.065 \\
&\te channels  & 9  & 10.6 & 0.30 \\
&\tb temporal  & 9  & 13.1 & 0.16 \\
&\tb channels  & 9  & 12.3 & 0.20 \\
\hline
\multirow{4}{*}{IT} 
&\te temporal  & 9  & 14.8 & 0.10 \\
&\te channels  & 9  & 2.42 & 0.98 \\
&\te temporal  & 9  & 7.55 & 0.58 \\
&\tb channels  & 9  & 13.5 & 0.14 \\ 
\enddata
\tablecomments{The first bin is not used in this analysis
since it can be contaminated by instrumental effects.
 }
\end{deluxetable}

\subsection{Propagation of instrumental uncertainties}
Additionally, we have modelled the potential systematic 
effects from mis-estimation of various instrumental 
characteristics using Monte Carlo simulations of
signal-only time ordered 
data, processed 
varying those parameters randomly over 
their range of uncertainty with a Gaussian distribution.
The parameters that have been changed are, the relative
calibration between channels ($\pm 0.8 \%$), the polarization 
efficiency ($\pm 0.03$), the bolometer time constants 
($\pm 10 \%$), the beam ($\pm 0.3'$), and
the angles of the polarizers axes respect to the telescope
frame ($\pm 2^\circ$). These ranges are the 
uncertainties on those instrumental parameters as
described in \cite{masi05}.

\begin{figure}[tb]
\includegraphics[angle=90,width=\columnwidth]{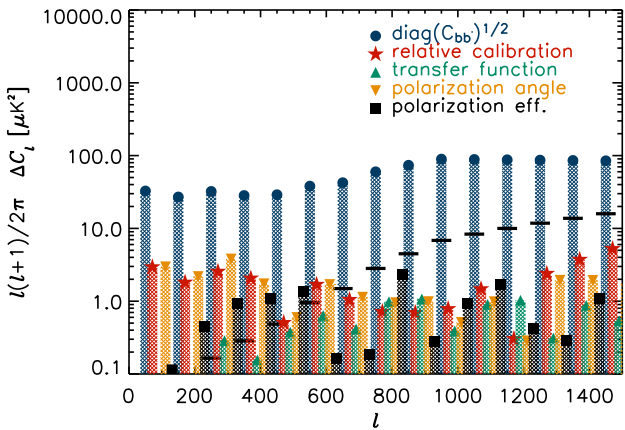}
\caption{
Propagation of instrumental uncertainties in the \te
error-bars. The dots are the square root of the 
diagonal part of the covariance matrix, 
relative calibration is varied by $\pm 0.8 \%$,
polarization efficiency by $\pm 0.03$, time constants 
of the transfer function 
by $\pm 10 \%$, the angles of the polarizers respect 
to the telescope
frame by $\pm 2^\circ$, and the beam 
(plotted as horizontal thicks) by $\pm 0.3'$.
The error-bars ($\Delta C_\ell$) generated
by uncertainties in instrumental characteristics are
one order of magnitude lower than the errors due
to noise and sampling variance (from NA pipeline here).
Those error-base are not treated as an increased error,
but rather as a systematic effect which is correlated
bin-to-bin, and marginalized over in the parameter estimation
as described in \citep{bridle2002}.
}
\label{fig:te_sys}
\end{figure}

As shown in Figure~\ref{fig:te_sys}, the potential errors 
from mis-estimation of these instrumental 
parameters are all at least one order of magnitude lower
than the statistical error-bars of the dataset.
The simulation uses the $\Lambda$CDM
model of~\cite{spergel03} fit to
WMAP (year~1), Acbar, and CBI.

\section{Conclusion}
\begin{figure}[tb]
\plotone{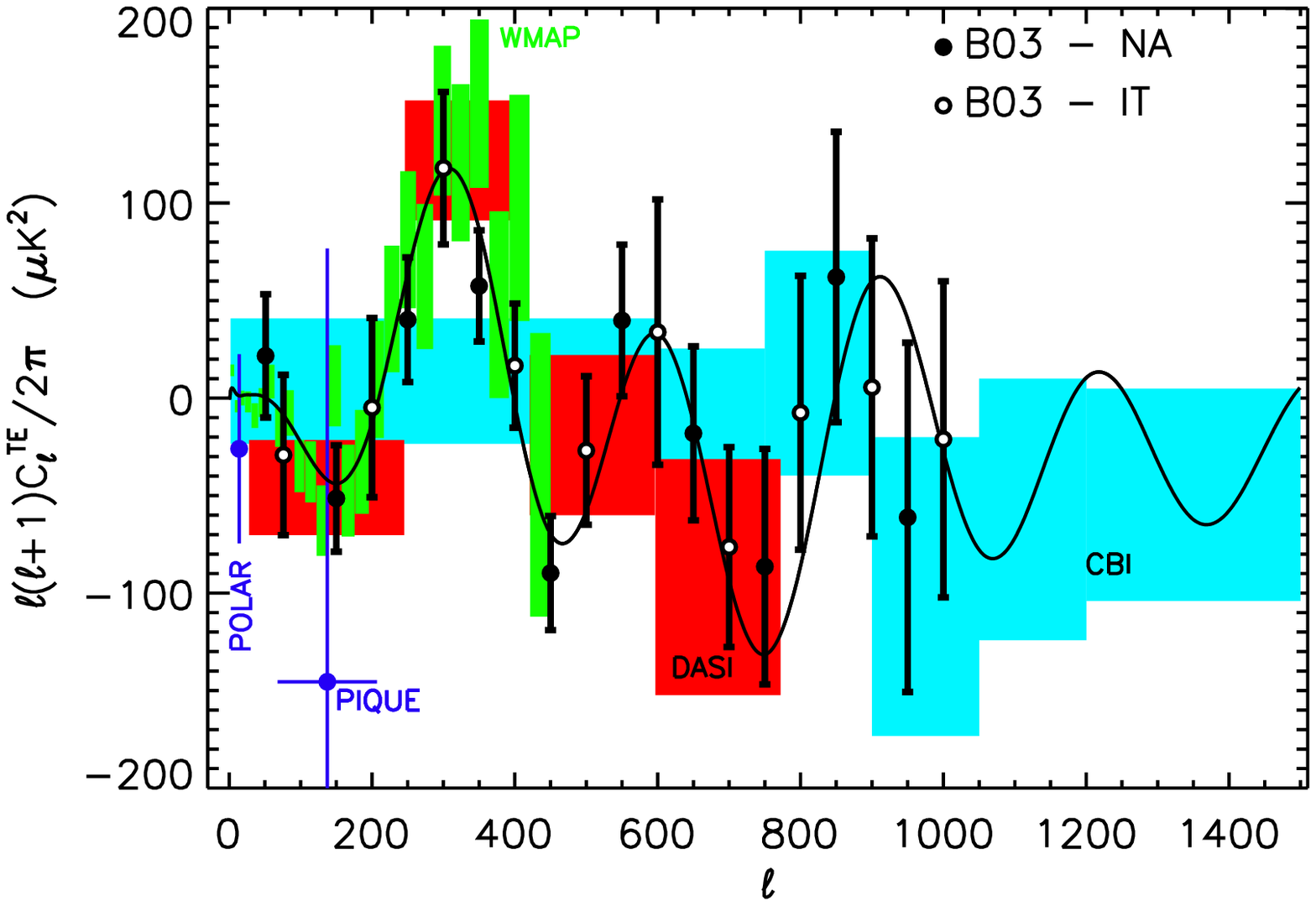}
\caption{
Collection of results \te power spectrum
from recent experiments. The \boom data are
NA and IT, with alternate binning. 
WMAP data are from \cite{kogut03}, CBI from \cite{cbi2004},
DASI from \cite{dasi2}, PIQUE from \cite{pique2003}, POLAR
from \cite{polar2003}.
}
\label{fig:comparison}
\end{figure}

We have detected
the presence of polarization of the \cmb with high statistical
significance (3.5~$\sigma$ combining the bins). This detection
of \te confirms and improves previous lower frequency 
detections \citep{kogut03,dasi2}
using a completely independent technology.
The robustness of these results against foreground contamination
effects is thus strengthened, and its cosmological origin
confirmed. 
A summary of all measurements of the \te
spectrum is shown in Figure~\ref{fig:comparison}.

The B03 \te data show a  2~$\sigma$ anti-correlation 
at large angular scales ($50 < \ell < 150$). This, as previously detected by
the WMAP experiment \citep{peiris}, is consistent with the presence of
superhorizon adiabatic fluctuations \citep{spergel97} and does not
support models based on active perturbations like topological defects
\citep{turok96}. In
active models the perturbations are continuously produced
by the causal field and lead to a positive correlation in the \te
spectrum. While cosmic string and textures models are already ruled
out by the presence of peaks in the \cmb temperature power spectrum,
active models may be constructed \citep[see e.g.][]{durrer} to
mimic the \TT data but not the \te spectrum.

\section{Acknowledgement}
We gratefully acknowledge support from
CIAR, CSA and NERSC in Canada,
ASI, University La Sapienza and PNRA in Italy,
PPARC and the Leverhulme Trust in the UK, and
NASA (awards NAG5-9251 and NAG5-12723) and
NSF (awards OPP-9980654 and OPP-0407592) in the USA.
Additional support for detector development was provided by CIT and JPL.
CBN acknowledges support from a Sloan Foundation Fellowship; WCJ and TEM
were partially supported by NASA GSRP Fellowships.
Field, logistical, and flight support was outstandingly
supplied by USAP and NSBF;  data recovery was especially appreciated.
This research used resources at NERSC, supported
by the DOE under Contract No. DE-AC03-76SF00098, and the MacKenzie
cluster at CITA, funded by the Canada Foundation for Innovation.
We also thank the CASPUR (Rome-ITALY) computational facilities
and the Applied Cluster Computing Technologies
Group at the Jet Propulsion Laboratory for computing
time and technical support.
Some of the results in this paper have been derived using
the HEALPix \citep{healpix} package and nearly all
have benefited from the FFTW3 implementation
of Fast Fourier Transform \citep{fftw}.  The \boom
field team is also grateful to the Coffee House at McMurdo
Station, Antarctica, for existing.

\bibliographystyle{apj}
\bibliography{te}

\end{document}